\documentclass[article]{aa}

\usepackage{graphicx}
\usepackage{amsmath}
\usepackage[varg]{txfonts}

\usepackage[dvipsnames]{xcolor}
\usepackage{hyperref}
\hypersetup{pdfborder = {0 0 0},
    colorlinks = true,
    linkbordercolor = {white},
    citecolor=NavyBlue,
    urlcolor=Green
}
\bibpunct{(}{)}{;}{a}{}{,} 

\newcommand{\orcidlink}[1]{\protect\href{https://orcid.org/#1}{\protect\includegraphics[width=8pt]{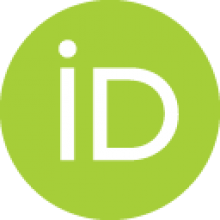}}}

\begin{document}

\title{Phakoseismology: \\ Asteroseismology of gravitationally microlensed stars}
\titlerunning{Asteroseismology of microlensed stars}
\authorrunning{Marcussen \& Lund}

\author{
Marcus L. Marcussen\inst{\ref{aarhus}}\corrauth{marcus@phys.au.dk}\orcidlink{0000-0003-2173-0689}
\and
Mikkel N. Lund\inst{\ref{aarhus}}\email{mikkelnl@phys.au.dk}\orcidlink{0000-0001-9214-5642}
}

\institute{
Department of Physics and Astronomy, Aarhus University, Ny Munkegade 120, 8000 Aarhus C, Denmark\label{aarhus}
}

\date{Received 21 August 2026 / Accepted 25 September 2026}

\abstract
% context
{Asteroseismology relies on disk-integrated photometry or radial velocities, in which geometric cancellation suppresses modes of high spherical degree ($l\geq4$). Observations are thus limited to deeply penetrating, low-degree modes, leaving the outer envelope inaccessible.}
% aims
{We investigate whether a high-magnification microlensing event can temporarily lift this limitation by differentially magnifying the surface of an oscillating source star.}
% methods
{We modelled the visibility of solar-like oscillations during microlensing, weighting the surface by limb darkening and the local magnification. We computed multiplet visibilities and the power distribution among azimuthal components, and built a likelihood model for recovering the 3D spin-axis orientation from resolved rotational multiplets.}
% results
{When the lens-source separation becomes comparable to the source's angular radius, the magnification gradient acts as a transient spatial filter that breaks the symmetry of disk integration, raising high-degree mode visibilities by up to three orders of magnitude. The same non-axisymmetric weighting makes the relative power of azimuthal components depend on both the inclination and sky-projected azimuth of the spin axis, imprinting the full spin orientation onto the power spectrum. With $5\%$ relative-power uncertainties, we recover the orientation to $\sigma_i=2.9^\circ$ and $\sigma_\gamma=11.5^\circ$.}
% conclusions
{Microlensing thus does more than amplify distant stars: its magnification pattern enables spatially selective asteroseismology. The effect is strongest for rare close encounters and unlikely to support large samples, and the newly visible high-degree modes will overlap with rotationally split and mixed modes, complicating mode identification. Nevertheless, individual high-quality events may open a window on envelope physics and spin geometries for stars in the Galactic bulge and beyond.}


\maketitle
\nolinenumbers

\section{Introduction}
\label{sec:intro}

Massive objects deflect light through their gravitational fields. When a foreground mass (the lens) passes close to the line of sight to a more distant star (the source), it produces multiple images separated on the sky by an angular scale set by the Einstein radius, $\theta_E$ \citep{Einstein1936}. For typical stellar-mass lenses in the Milky Way, these images are unresolved, so the event is observed as an achromatic, time-variable magnification of the background source, described to first order by the Paczy\'nski light curve \citep{Paczynski1986}.

Because the magnification depends only on gravity, microlensing is a powerful probe of both the lens and the source. Deviations from the single-lens light curve reveal structure in the lens system and have made microlensing a leading method for discovering and characterising exoplanets \citep{Gaudi2012}.  During high-magnification events, finite-source effects imprint spatial information about the source, enabling measurements of surface properties such as limb darkening \citep{Albrow1999}. In this work, we propose that the same differential magnification can reveal oscillation signatures that are otherwise hidden in unresolved starlight.

Solar-like oscillations are standing acoustic waves excited by near-surface convection and trapped within the stellar interior \citep{Chaplin2013}. Their resonant frequencies encode the internal sound-speed profile and thereby provide uniquely precise constraints on fundamental stellar properties such as masses, radii, and ages \citep{Lebreton2014,Buchele2025}, as well as on internal rotation and mixing \citep{Aerts2021}. For ensembles of red giants, these capabilities underpin Galactic archaeology \citep{Miglio2013}. The information content of a mode depends on its spherical degree, $l$: low-degree modes penetrate deeply into the star, whereas high-degree modes have shallower lower turning points and are increasingly sensitive to the outer envelope.

The fundamental limitation is spatial resolution. In seismic studies of the Sun (helioseismology), the surface is spatially resolved, so observations of high-degree modes are possible and enable tight constraints to be placed on envelope structure, including helium-ionisation signatures, convection-zone boundaries, and near-surface rotation \citep{Dalsgaard2002,Houdek2007,Basu2009}. For other stars, however, the surface is unresolved, and the contributions from different parts of the disk cancel out when integrated: the bright and dim patches of a high-degree mode average out in photometry, just as its blue- and redshifted regions average out in the disk-integrated velocity. Either way, high-degree modes are rendered undetectable, and asteroseismic inference is therefore largely confined to low-degree modes \citep[$l\leq3$;][]{Lund2014,Lund2017}, while the outer envelope, where high-$l$ modes primarily propagate, remains observationally inaccessible.

So far, microlensing and asteroseismology have been combined only in limited ways, mainly through the total amplification of the source. \citet{Li2019} used the microlensed red giant \object{OGLE-2017-BLG-1186} to detect solar-like oscillations in a star at a distance of approximately $11.5\,$kpc and infer its global stellar properties. With the large sample of microlensing events expected from the Galactic Bulge Time Domain Survey of the \textit{Nancy Grace Roman Space Telescope} \citep{Penny2019, Yee2023}, such measurements may become possible for a much larger number of distant stars \citep{Gould2015,Weiss2025}. \citet{Sajadian2020,Sajadian2020B} explored a complementary synergy between microlensing and asteroseismology, studying how radial and non-radial pulsations modify microlensing light curves. Here we ask the reverse question: how does microlensing change the observability of the oscillations themselves?

\begin{figure}
\centering
\includegraphics[width=\columnwidth]{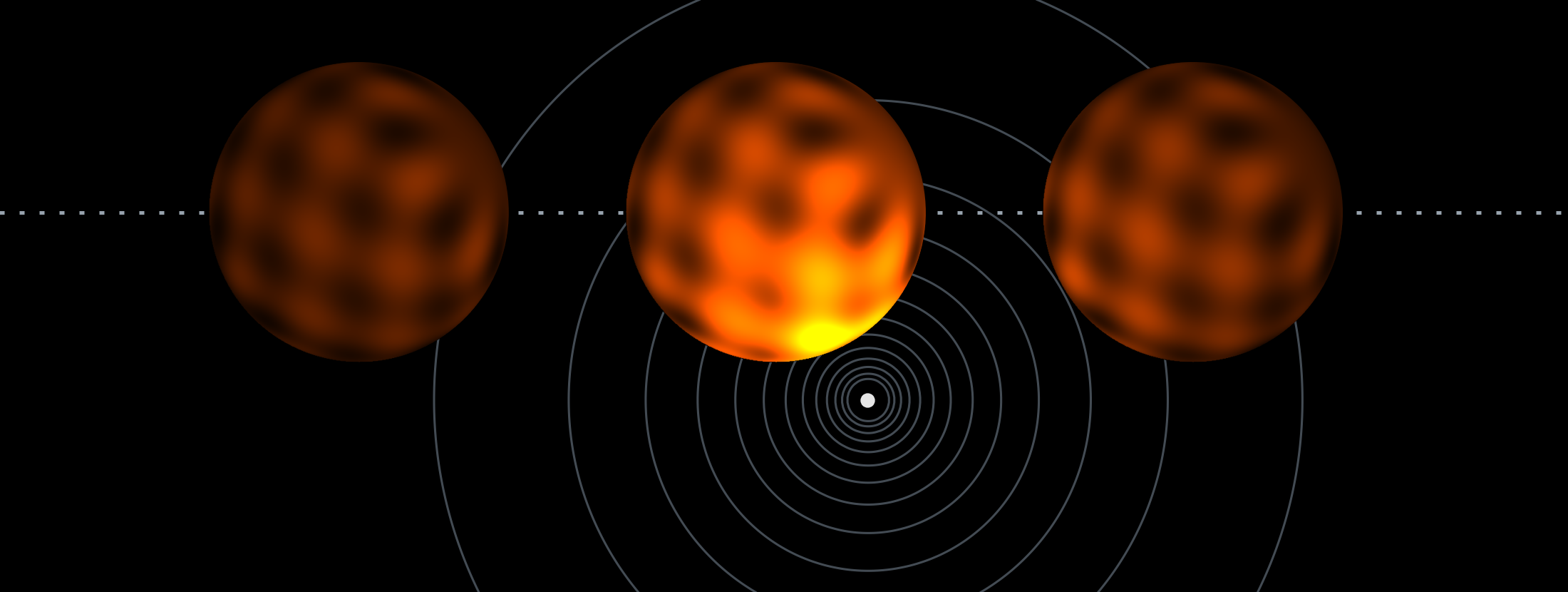}
\caption{Microlensed oscillating source. The three coloured disks show the source star at different positions as its line of sight passes close to that of a lensing object. The gray dot shows the projected position of the lens, and the thin lines correspond to equal lens magnification strengths. Exaggerated bright and dark surface regions on the disks illustrate the local flux perturbations caused by stellar oscillations. As the lens and source approach each other in projection, the magnification varies across the stellar disk and weights different parts of the mode pattern unequally. Even when the disks overlap in projection, the source is not occulted. Light is deflected to angles of order the Einstein radius, far beyond the physical disk of the lens, so close approaches produce magnification rather than shadowing.}
\label{fig:illustration}
\end{figure}

We investigate what we refer to as `phakoseismology', the asteroseismology of gravitationally microlensed stars (from the Greek phakos, meaning lens). The central idea is that a close microlensing event imposes a time-dependent, non-axisymmetric weighting on the stellar disk (Fig.~\ref{fig:illustration}). Instead of merely increasing the total flux, the lens magnifies different parts of the oscillation pattern by different amounts. This both eliminates the geometric cancellation that normally hides high-degree modes and changes how power is distributed among the azimuthal components of rotationally split multiplets. The latter effect imprints information about the full three-dimensional spin orientation onto the oscillation power spectrum, rather than only the inclination to the line of sight. To our knowledge, this use of the magnification pattern as a spatially selective seismic filter is new and complementary to earlier work.

The paper is organised as follows. Section~\ref{sec:methods} describes the lensing geometry, the visibility calculation, and the likelihood model for spin-axis recovery. Section~\ref{sec:results} presents the effects of microlensing on multiplet visibilities and azimuthal power distributions. Section~\ref{sec:discussion} discusses the prospects and limitations of the method, and Sect.~\ref{sec:conclusion} summarises the main conclusions.

\section{Methods}
\label{sec:methods}

\subsection{Geometry and coordinate systems}
\label{ssec:geom}

\begin{figure*}
\centering
\includegraphics[width=\textwidth]{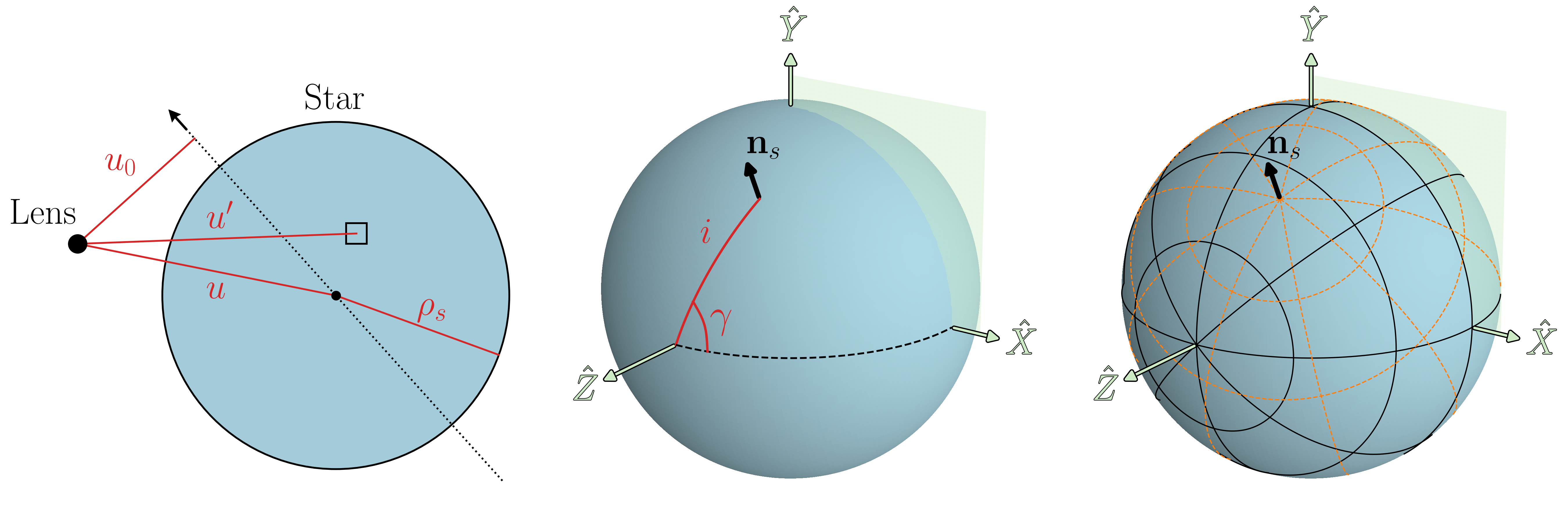}
\caption{Geometry and coordinate systems. \textit{Left:} Sky-plane geometry of a close microlensing event, with all angular separations scaled to the Einstein radius. The source is separated from the lens by $u$ and has angular radius $\rho_s$; the nearest approach occurs at $u_0$, and the surface is discretised into elements at separation $u'$ from the lens. \textit{Middle:} Orientation of the stellar rotation axis, $\mathbf{n}_s$. The coordinate system has $\hat{Z}$ pointing towards the observer, so the $\hat{X}$--$\hat{Y}$ plane defines the sky plane (shown transparent). The inclination ($i$) and azimuth ($\gamma$) define the three-dimensional orientation of the spin axis. \textit{Right:} The two coordinate systems used in this work. The observer's frame is shown as solid black curves at fixed polar angle $\theta$ and azimuth $\phi$; the stellar inertial frame is shown as dashed orange curves.}
\label{fig:geometry}
\end{figure*}

We describe brightness variations across the stellar surface using two right-handed spherical coordinate systems (see the right panel of Fig.~\ref{fig:geometry}). The stellar inertial frame, $(r',\theta',\phi')$, is defined with the stellar rotation axis aligned to $\theta'=0$. The observer's frame, $(r,\theta,\phi)$, is defined with the line of sight at $\theta=0$, so the visible disk spans $0\leq\theta\leq\pi/2$.

The two frames are related by a rigid rotation. We use $i\in[0,\pi]$ to denote the inclination of the stellar rotation axis relative to the line of sight, and $\gamma\in[0,2\pi)$ for the sky-projected azimuthal angle of the spin axis (see the middle panel of Fig.~\ref{fig:geometry}). If $\mathbf{r}$ and $\mathbf{r}'$ are the Cartesian position vectors of a surface element in the observer's frame and the stellar inertial frame, respectively, then $\mathbf{r}' = R(i,\gamma)\,\mathbf{r}$, where
\begin{equation}
R(i,\gamma)=
\begin{pmatrix}
\cos\gamma\cos i & \sin\gamma\cos i & -\sin i\\
-\sin\gamma       & \cos\gamma       &  0      \\
\cos\gamma\sin i & \sin\gamma\sin i &  \cos i
\end{pmatrix}.
\label{eq:Rmatrix}
\end{equation}
Oscillation eigenfunctions are defined in the stellar frame using spherical harmonics $Y_l^m(\theta',\phi')$, while flux integrals are evaluated over the visible disk in the observer's frame. Throughout the paper, $n$, $l$, and $m$ denote the radial order, spherical degree, and azimuthal order.

\subsection{Microlensing magnification}
\label{ssec:magnification}

We scaled all angular separations by the angular Einstein radius \citep{Einstein1936},
\begin{equation}
\theta_E=\sqrt{\frac{4GM}{c^2}\frac{D_S-D_L}{D_LD_S}},
\label{eq:einstein_radius}
\end{equation}
where $G$ is the gravitational constant, $c$ is the speed of light, $M$ is the lens mass, and $D_L$ and $D_S$ are the lens and source distances. The scaled projected source-lens separation is $u\equiv\theta/\theta_E$, and the scaled angular radius of the source is $\rho_s\equiv\theta_s/\theta_E$.

For a point source and a point lens, the magnification is \citep{Paczynski1986}
\begin{equation}
A(u) = \frac{u^2 + 2}{u\sqrt{u^2 + 4}}.
\label{eq:point_magnification}
\end{equation}
For an extended source, the relevant separation differs across the stellar disk. We therefore discretised the surface into elements at projected position $(r,\phi)$ relative to the source centre. A surface element at colatitude $\theta$ has a projected distance of
\begin{equation}
r(\theta)=\rho_s\sin\theta
\end{equation}
from the source centre, and its instantaneous scaled separation from the lens is \citep{WittMao1994}
\begin{equation}
u'(\theta,\phi;u)=
\sqrt{u^2+r(\theta)^2-2\,u\,r(\theta)\cos\phi},
\label{eq:uprime_dimless}
\end{equation}
The geometry is shown in the left panel of Fig.~\ref{fig:geometry}. This expression assumes that the lens lies on the positive $x$-axis of the sky-plane coordinate system at the epoch being modelled; a general lens trajectory can be represented by rotating this coordinate system.

\subsection{Mode visibility}
\label{ssec:visibility}

We modelled the disk-integrated intensity fluctuations, $I(t)$, as the surface integral of an oscillation pattern weighted by the function $W(\theta,\phi)$, which combines limb darkening and microlensing magnification. Following \citet{Gizon2003}, the signal can be written as
\begin{equation}
I(t) =
\mathrm{Re}
\sum_{n l m}
A_{nl}\, e^{it(\omega_{ln}-m\Omega)+i\phi_{nlm}}\;
V_{lm},
\label{eq:intensity_signal}
\end{equation}
where $A_{nl}$ is the intrinsic mode amplitude, $\omega_{ln}$ is the angular frequency in the absence of rotation, $\Omega$ is the angular rotation rate, and $\phi_{nlm}$ is a random phase. The visibility factor encodes the geometric cancellation and the observing geometry,
\begin{equation}
V_{lm}(i,\gamma;u)=
\int_{\mathrm{disk}}
W(\theta,\phi;u)\;
Y_l^m\!\left(R_{i,\gamma}(\theta,\phi)\right)
\cos\theta\,d\Omega,
\label{eq:visibility_factors}
\end{equation}
The intrinsic amplitudes ($A_{nl}$) contain the excitation and damping physics, which we did not model. The weighting function is the product of the point-lens magnification of each surface element and the limb-darkening law,
\begin{equation}
W(\theta,\phi;u)=A\!\left(u'(\theta,\phi;u)\right)\,
L(\cos\theta).
\label{eq:W_norm}
\end{equation}
We adopted a quadratic limb-darkening law,
\begin{equation}
L(\cos\theta)=
1-c_1(1-\cos\theta)-c_2(1-\cos\theta)^2,
\label{eq:LD_unscaled}
\end{equation}
where $c_1$ and $c_2$ are fixed limb-darkening coefficients for the simulated passband. We use $c_1$ and $c_2$ rather than the more common $u_1$ and $u_2$ to avoid confusion with the source-lens separation $u$.

The total squared visibility of an angular multiplet is
\begin{equation}
V_l^2(i,\gamma;u)=
\sum_{m=-l}^{l} |V_{lm}(i,\gamma;u)|^2,
\label{eq:multiplet_visibility}
\end{equation}
and we compared multiplets using the visibility relative to the radial mode,
\begin{equation}
\frac{V_l^2}{V_0^2}=
\frac{\sum_{m=-l}^{l}|V_{lm}(i,\gamma;u)|^2}
{|V_{00}(i,\gamma;u)|^2}.
\label{eq:multiplet_power}
\end{equation}
The relative power in an azimuthal order $m$ within a multiplet is
\begin{equation}
p_{lm}(i,\gamma\mid u)=
\frac{|V_{lm}(i,\gamma;u)|^2}
{\sum_{m'=-l}^{l}|V_{lm'}(i,\gamma;u)|^2}.
\label{eq:azimuth_order_power}
\end{equation}
For an axisymmetric weighting function, such as limb darkening without microlensing, the dependence on the sky-projected azimuth ($\gamma$) vanishes, and the relative powers reduce to the classical inclination functions for unresolved stars \citep{Gizon2003}. Microlensing changes this because $W(\theta,\phi;u)$ is generally non-axisymmetric during a close event.

\subsection{Likelihood analysis for spin orientation}
\label{subsec:orientation}

We used the relative powers of resolved azimuthal components to quantify the information available on the stellar spin orientation. At a given source-lens separation $u$, the model predicts $p_{lm}(i,\gamma\mid u)$ from Eq.~\eqref{eq:azimuth_order_power}. We assumed that each measured relative power $p^{\rm obs}_{lm}(u)$ has an independent Gaussian uncertainty $\sigma$. For one multiplet of degree $l$, the likelihood is
\begin{equation}
\mathcal{L}_{l}(i,\gamma \mid u)=
\prod_{m=-l}^{l}
\frac{1}{\sqrt{2\pi}\,\sigma}
\exp\!\left[
-\frac{\left(p^{\rm obs}_{lm}(u)-p_{lm}(i,\gamma\mid u)\right)^2}
{2\sigma^2}
\right].
\label{eq:likelihood}
\end{equation}
The total likelihood is the product over all observed multiplets and all observational epochs (separations $u$),
\begin{equation}
\mathcal{L}(i,\gamma)=
\prod_{u'}\prod_{l'} \mathcal{L}_{l'}(i,\gamma\mid u').
\label{eq:total_likelihood}
\end{equation}
This calculation is not a full detectability analysis. It isolates the geometric information that would be available if the azimuthal components of rotationally split multiplets were resolved and their relative powers could be measured.

\section{Results}
\label{sec:results}
\begin{figure*}
\centering
\includegraphics[width=\textwidth]{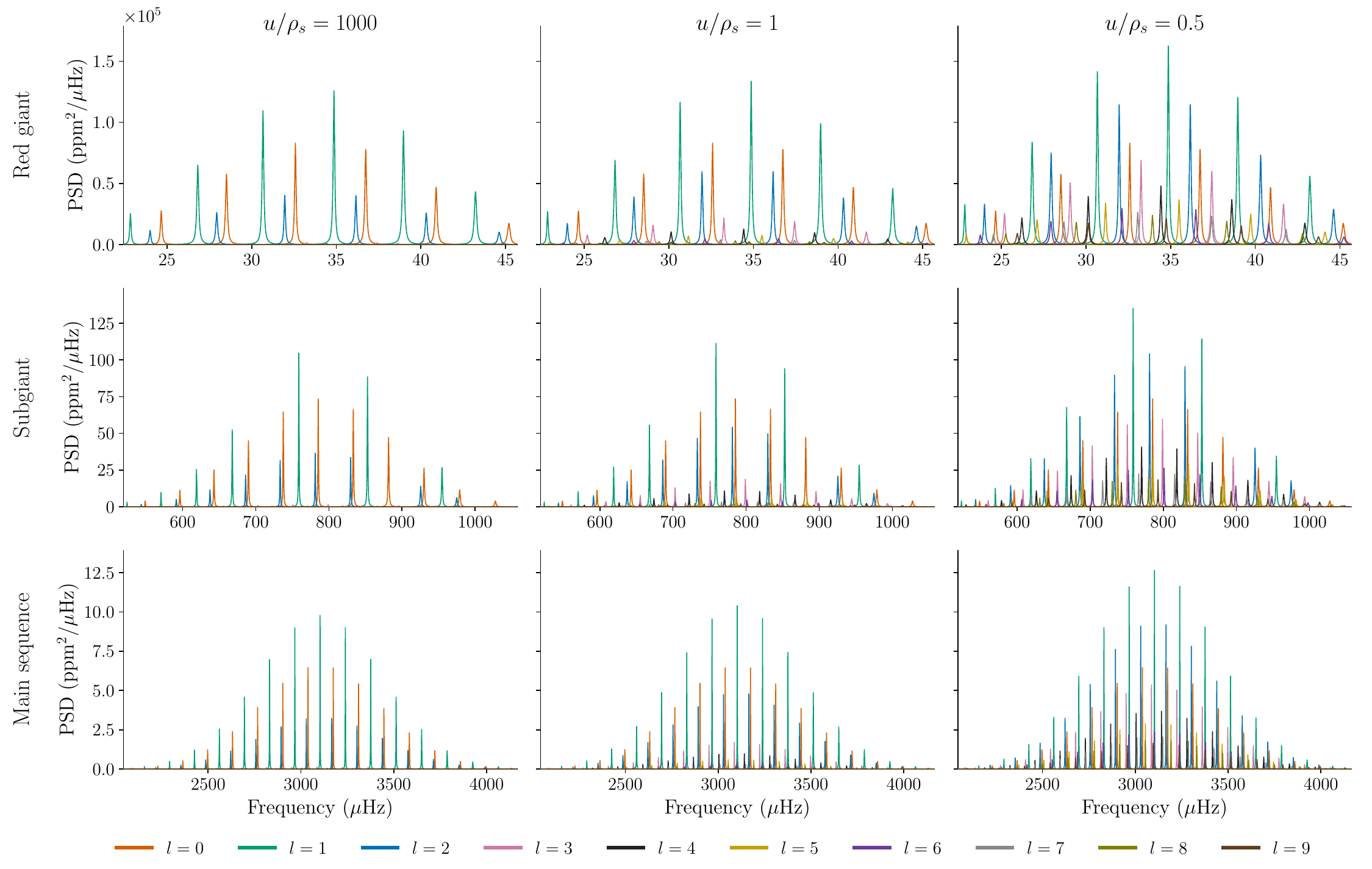}
\caption{Illustrative noise-free oscillation power density spectra of a $1\,M_\odot$ star at three evolutionary stages. The spectra include modes up to $l=9$ and use the microlensing-modified mode visibilities calculated in this work. Mixed modes, instrumental noise, and effects from rotational splitting and granulation have been omitted. Columns from left to right show source-lens separations of $u/\rho_s=1000$, $1$, and $0.5$, while rows from top to bottom show the red giant, subgiant, and main-sequence stages.}
\label{fig:power_spectra}
\end{figure*}

\subsection{Microlensing increases high-degree mode visibility}
\label{ssec:visibility_results}
\begin{figure}
\centering
\includegraphics[width=\columnwidth]{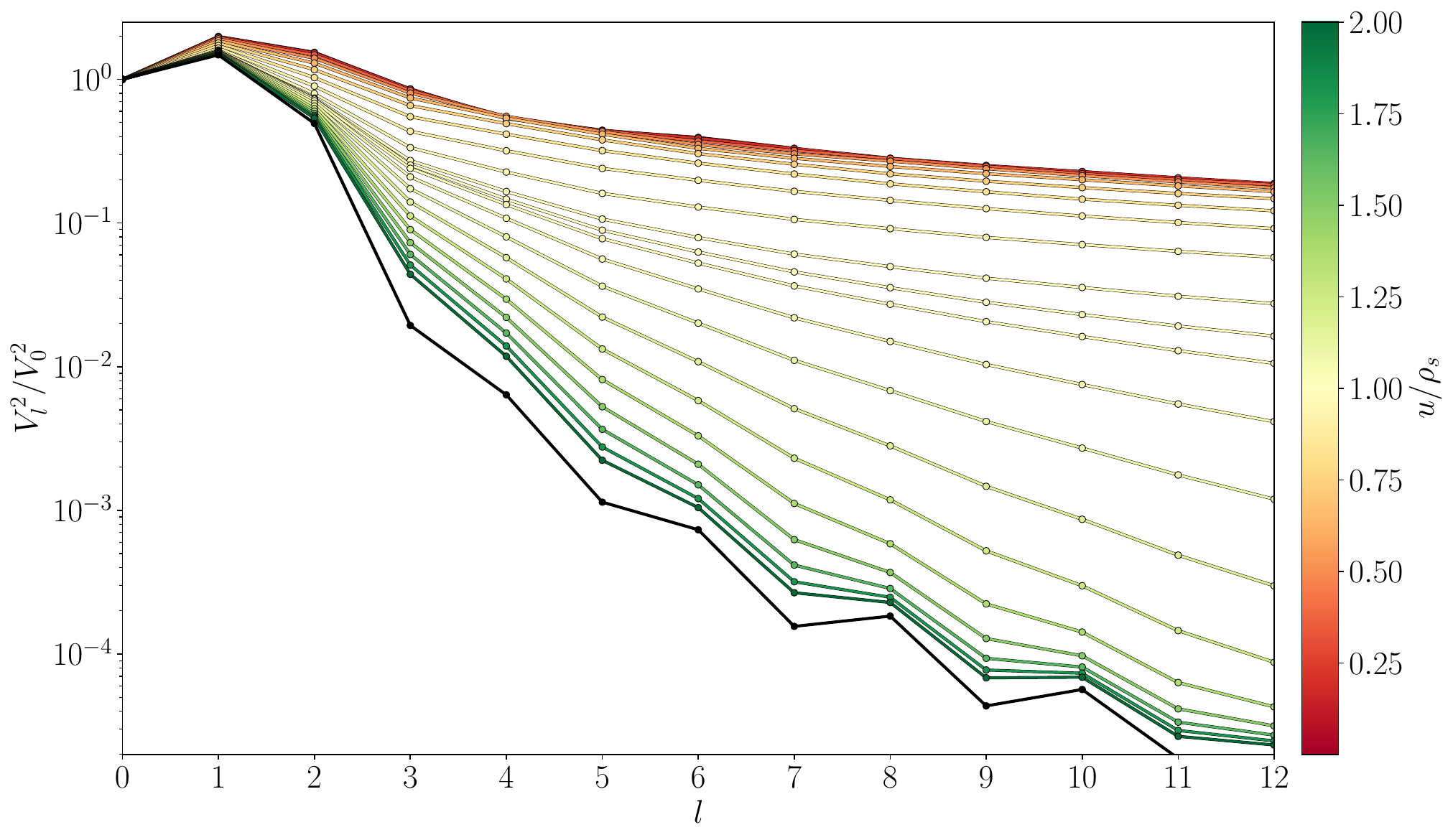}
\caption{Microlensing enhances the visibility of high-degree modes. Relative squared visibilities $V_l^2/V_0^2$ are shown for angular degrees $l=0$--$12$, coloured by the scaled source-lens separation $u$, which varies from $0$ (central transit) to $2\rho_s$. Note the orders-of-magnitude increase in visibility for high-$l$ modes as the lens approaches the stellar disk ($u\lesssim\rho_s$).}
\label{fig:multiplet_power}
\end{figure}

We first computed the bolometric squared visibility (i.e. power), normalised to the radial mode, $V_l^2/V_0^2$, for modes with $l=0$--$12$. For the fiducial case, shown in Fig.~\ref{fig:multiplet_power}, the source radius and lens mass were set to $1\,R_\odot$ and $1\,M_\odot$, respectively, and the source and lens distances were set to $8\,$kpc and $4\,$kpc, respectively.

Figure~\ref{fig:multiplet_power} shows that far from the lens, where magnification is nearly constant across the source, visibility declines rapidly with increasing $l$, rendering modes with $l>3$ undetectable. But once the projected separation approaches the stellar radius ($u \sim \rho_s$), the visibilities of high-degree modes rise by orders of magnitude. At $l=8$, for example, the visibility increases by a factor of ${\sim}10^3$ relative to the unlensed case, making the mode as visible as an unlensed quadrupole $l=2$ mode. The lens effectively acts as a spatial filter that transforms the unresolved star into a partially resolved object, making the detection of more envelope-sensitive modes feasible.

The increase is strongest for high $l$, but it does not remove the overall decline of visibility with degree. Microlensing therefore does not make all high-degree modes equally observable. Instead, it raises the power of some modes that would normally be hidden into a range where detection may be possible for favourable sources and observing conditions.

Illustrations of how mode visibility translates to the frequency domain are shown in Fig.~\ref{fig:power_spectra}. Here, representative oscillation limit spectra (noise free) are shown for a $1\,M_\odot$ red giant ($10\,R_\odot$), subgiant ($2\,R_\odot$), and main-sequence star ($1\,R_\odot$), including modes up to $l=9$. Mode frequencies were calculated from representative stellar models using the Aarhus adiabatic oscillation package (ADIPLS; \citealt{Christensen-Dalsgaard2008}). The evolved models include only the nominal acoustic mode at each radial order and degree. The full mixed-mode spectrum would be considerably denser, and together with rotational splitting, granulation, instrumental noise, and the finite, time-dependent observing window of the event, this would complicate both mode identification and the measurement of individual mode heights. The spectra are constructed from Lorentzian profiles with bolometric mode amplitudes from \citet{Corsaro2013}, and a Gaussian oscillation-power envelope following \citet{Mosser2012}. Within each stellar spectrum, the mode linewidth is held constant and prescribed using \citet{Vrard2018} for the red giant and \citet{Appourchaux2012} for the subgiant and main-sequence star.

\subsection{Microlensing encodes the stellar spin orientation}
\label{ssec:spin_results}

The orientation of a star's rotation axis in space is fully described by two angles (Fig.~\ref{fig:geometry}): the inclination $i$ between the rotation axis and the line of sight, and the sky-projected azimuth ($\gamma$), which specifies the direction in which the axis is tilted on the sky plane relative to the microlensing trajectory. In standard disk-integrated asteroseismology, the rotational splitting of oscillation frequencies allows $i$ to be measured \citep{Gizon2003,Dziembowski1977}, because the relative power among the split components depends on how the mode pattern projects onto the line of sight. The azimuth ($\gamma$), however, remains entirely unconstrained: rotating the spin axis around the line of sight leaves the axisymmetric disk weighting unchanged, so all azimuthal orientations produce identical power spectra.

The same spatial selectivity that lifts geometric cancellation also breaks this degeneracy. Once the lens imposes a non-axisymmetric weighting across the disk, the observed power depends on how the source-lens trajectory is oriented relative to the stellar equator. We therefore examined how power is redistributed among the azimuthal orders $m$ of a given multiplet.

\begin{figure*}
\centering
\includegraphics[width=\textwidth]{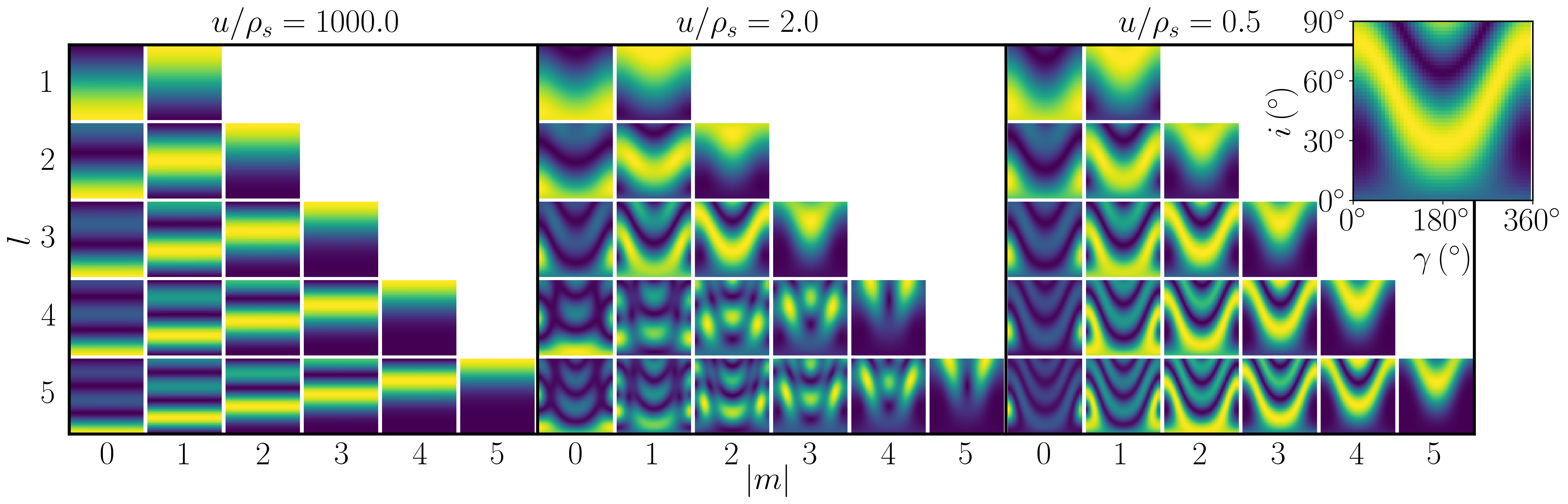}
\caption{Azimuthal power distribution reveals the spin orientation. The colour map represents the relative distribution of power among azimuthal orders $|m|$ for a given multiplet $l$, plotted as a function of the spin-axis orientation, with inclination ($i$) on the vertical axis and azimuth ($\gamma$) on the horizontal axis. From left to right, the lens moves from a large separation ($u=1000\,\rho_s$) to a close transit ($u=0.5\,\rho_s$). At close separation, the distribution of power among the azimuthal orders depends on the angle $\gamma$, breaking the rotational degeneracy.}
\label{fig:m_power}
\end{figure*}

Figure~\ref{fig:m_power} shows the relative power distribution among the azimuthal orders for different spin orientations and source-lens separations. At $u=1000\,\rho_s$, the lens is effectively far from the source; the distribution is symmetric and independent of $\gamma$, and we reproduce the results of \citet{Gizon2003}. As the lens crosses the disk ($u=2\,\rho_s$ and $u=0.5\,\rho_s$), the symmetry breaks: the relative powers of the $|m|$ components now depend uniquely on the three-dimensional orientation of the spin axis relative to the lensing trajectory.

\begin{figure*}
\centering
\includegraphics[width=\textwidth]{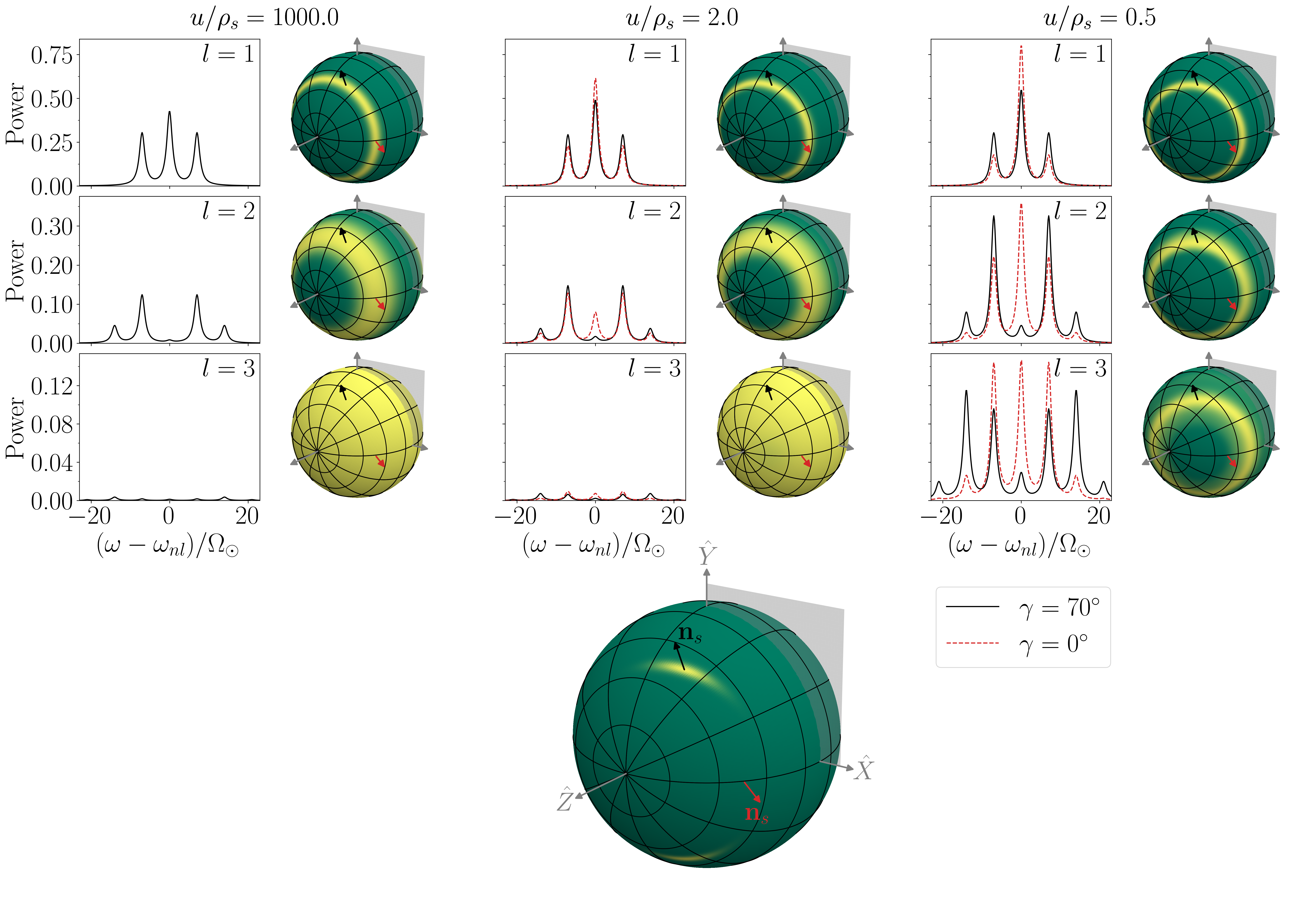}
\caption{Recovery of the three-dimensional stellar spin axis. We present likelihood maps for the inclination ($i$) and sky-projected azimuth ($\gamma$) derived from the spectral power of multiplets $l=1,2,3$ (each displayed on a separate row) at three lens separations $u/\rho_s=1000,\,2,\,0.5$ (along each column). The input orientation $(i,\gamma)=(50^\circ,70^\circ)$ and the reference orientation $(i,\gamma)=(50^\circ,0^\circ)$ are shown in black and red, respectively. The spheres follow the format of Fig.~\ref{fig:geometry} and indicate the stellar spin direction, while the colour map shows the posterior probability distribution given the input orientation. Individual modes or epochs leave extended allowed regions, while the combined likelihood (bottom panel) pinpoints the true orientation.}
\label{fig:orientation}
\end{figure*}

Sampling the power spectrum at multiple epochs during the event makes it possible to constrain the spin orientation. We tested this with synthetic observations of multiplets $l=1,2,3$ at three epochs ($u/\rho_s=1000,\,3,\,0.5$) for a star with inclination $i=50^\circ$ and azimuth $\gamma=70^\circ$, assuming independent uncertainties of $\sigma=5\%$ on the measured relative powers. As shown in Fig.~\ref{fig:orientation}, individual multiplets and individual epochs leave extended degeneracies, but the combination of the three snapshots (for three non-radial modes, i.e., nine measurements in total) tightly constrains the posterior and recovers the full $(i,\gamma)$ vector. For the assumed uncertainties of relative power, the inclination is recovered to within $\sigma_i=2.9^\circ$ and the azimuth to within $\sigma_\gamma=11.5^\circ$.

\subsection{Expected event rates}
\label{ssec:event_results}

To estimate the rate of phakoseismic events, we generated a synthetic
population of source and lens stars using the Besan\c{c}on Galactic
Model \citep{Robin2003}, requiring $D_L<D_S$ and drawing the impact
parameter $u_0$ uniformly from $\mathcal{U}(0,1)$. We classified an event
as phakoseismic when the lens passed within two source radii,
$u_0\leq2\rho_s$. This gives around $350$ phakoseismic events per year
across the Galactic bulge.

We then estimated how many of these events would be accessible to the
Galactic Bulge Time Domain Survey \citep{Yee2023} of the \textit{Nancy
Grace Roman Space Telescope}. Most of the sources are too faint for
precise asteroseismic measurements, and the survey covers only part of
the bulge. To account for these limitations, we applied a magnitude cut
of $H_{\rm AB}<17$ and normalised the number of $|u_0|<1$ microlensing
events to the expected mission yield, $N_{\rm tot}\simeq27{,}000$
\citep{Penny2019}. We estimate approximately two high-quality candidate
events to be detected per year in the \textit{Roman} field. About half are expected to
be red giants, whose large angular radii both increase $\rho_s$ and
extend the interval over which differential magnification is strong.

\section{Discussion}
\label{sec:discussion}

\subsection{Information in high-degree modes}
\label{ssec:science_discussion}

Microlensing can make unresolved stars temporarily behave as partially resolved seismic targets. The lens does not resolve the stellar disk directly, but it imposes a spatially varying magnification pattern across the surface. This pattern acts as a transient spatial filter, suppressing some parts of an oscillation pattern and enhancing others, and thereby reducing the cancellation that normally hides high-degree modes in disk-integrated observations.

The main scientific interest lies in the outer stellar envelope. High-degree acoustic modes have lower turning points at higher fractional radius and are therefore increasingly dominated by the properties of the near-surface layers \citep{Bellinger2017,Bellinger2020}. In the Sun, such modes strongly constrain helium-ionisation signatures, convection-zone structure, and near-surface rotation \citep{Dalsgaard2002,Basu2009}. In other stars, access to even a small number of higher-degree modes could provide an independent constraint on the surface term and on the envelope physics that currently limit the precision of asteroseismic ages and masses \citep{Ball2017,Mosumgaard2020,Buchele2024}.

The method is most promising for red giants. Their large radii increase the probability that the lens passes within a few source radii, and their oscillation amplitudes are larger than those of main-sequence solar-like oscillators. Red giants are also scientifically valuable because their seismic masses and ages are widely used in Galactic studies. A successful phakoseismic event would therefore provide the first spatially selective asteroseismic measurement of a star other than the Sun, offering a critical calibration point for envelope-sensitive stellar-structure theory that is otherwise attainable only for the Sun.

These prospects come with a caveat: the enhanced visibility of high-degree modes also makes the power spectrum more crowded, turning mode identification into a significant observational challenge. This is particularly true for evolved stars with rotationally split multiplets and dense mixed-mode spectra, and it is exacerbated by the limited frequency resolution of the lensing window.

\subsection{Spin orientation}
\label{ssec:spin_discussion}

The microlensing signal also carries information about the spin-axis orientation. In standard asteroseismology, the relative heights of rotationally split components constrain only the inclination $i$ \citep{Lund2014b,Benomar2014,vanEylen2014,Campante2016}; the sky-projected azimuth ($\gamma$) is inaccessible because the unresolved stellar disk is axisymmetric. During a close microlensing event, the power spectrum not only depends on the orientation of the stellar spin axis relative to the observer, but also on the orientation of the stellar equator relative to the lens trajectory.

This effect may be useful for systems in which a projected orbital direction is known. If the stellar spin axis can be measured in three dimensions, the true stellar obliquity can be inferred for microlensed binaries or planet-hosting systems \citep{Albrecht2022}, providing a complementary route to studying spin-orbit architectures in distant stellar populations. Such an application requires a favourable event, resolved rotational splitting, and accurate mode-height measurements.

\subsection{Time window and frequency resolution}
\label{ssec:timescale_discussion}

The main practical limitation is the short duration of the favourable alignment. The high-degree visibility increases only while the source-lens separation is comparable to the angular radius of the source. If we define the useful interval as $u(t)\leq2\rho_s$, its duration is
\begin{equation}
\Delta t_{\rm vis}=\frac{2\sqrt{(2\theta_s)^2-u_0^2\theta_E^2}}{\mu_{\rm rel}},
\label{eq:visibility_window_mu}
\end{equation}
where $\mu_{\rm rel}$ is the relative proper motion. Equivalently, in terms of the transverse relative velocity $v_{\rm rel}$ at the source distance,
\begin{equation}
\Delta t_{\rm vis}=\frac{2D_S\sqrt{(2\theta_s)^2-u_0^2\theta_E^2}}{v_{\rm rel}}.
\label{eq:visibility_window}
\end{equation}
For a nearly central event, this reduces to
\begin{equation}
\Delta t_{\rm vis}
\simeq
\frac{4R_s}{v_{\rm rel}}=32.2\,{\rm d}\,
\left(\frac{R_s}{R_\odot}\right)
\left(\frac{v_{\rm rel}}{{\rm km\,s}^{-1}}\right)^{-1}.
\label{eq:visibility_window_central}
\end{equation}
The corresponding Fourier resolution is approximately
\begin{equation}
\delta\nu
\simeq
\frac{1}{\Delta t_{\rm vis}}=\frac{v_{\rm rel}}{4R_s}
\simeq
0.36\,\mu{\rm Hz}\,
\left(\frac{v_{\rm rel}}{{\rm km\,s}^{-1}}\right)
\left(\frac{R_\odot}{R_s}\right).
\label{eq:frequency_resolution}
\end{equation}

For a red giant, the useful interval can last days to weeks, and the large radius also increases the finite-source probability. However, the longer oscillation periods of red giants partly offset this advantage. Fewer oscillation cycles are observed during the microlensing window, making individual mode frequencies and their temporal evolution more difficult to measure despite the longer duration of the event.

\subsection{Observational prospects}
\label{ssec:prospects_discussion}

Close finite-source encounters occur regularly across the Galactic
bulge (Sect.~\ref{ssec:event_results}).
Exploiting them, however, calls for a survey that combines
asteroseismic photometric precision on faint bulge stars with a field
of view wide enough to monitor a substantial fraction of the bulge --
a combination no existing facility provides. The \textit{Roman}
Galactic Bulge Time Domain Survey comes closest, and under its
observing conditions we estimate a few high-quality candidate events
per year. A complete assessment would need to include, for instance, the \textit{Roman} cadence, photometric uncertainties, blending, finite
exposure times, and the time-dependent lensed light curve.

\subsection{Other pulsators}
\label{ssec:pulsators_discussion}

The modification of the geometric cancellation during a microlensing
event may also be useful for classical pulsators, where mode
identification is often a central challenge \citep{Goupil2005,Kurtz2022,Bedding2023}. In this regard, $\delta$\,Sct stars are a
particularly interesting case because they can show many non-radial
modes of different spherical degrees, together with rotationally split
components, and the observed frequencies do not in general determine
the corresponding $(l,m)$ values uniquely. Multi-colour photometry is
widely used to provide additional mode-identification constraints
\citep{Watson1988,Balona1999}, with spectroscopy offering further
information \citep{Aerts2010}.

Phakoseismology could provide a different constraint by making the mode
visibility time-dependent. As the lens moves across the stellar disk,
modes with different surface geometries are weighted differently, so
their observed amplitudes should change in different ways during the event.
A benefit of classical pulsators over solar-like oscillators is their much higher (often at the millimagnitude level) photometric amplitudes, and their short periods (hours for $\delta$\,Sct stars). This means many pulsation cycles fit within even a brief lensing window, mitigating the frequency-resolution limitation
discussed in Sect.~\ref{ssec:timescale_discussion}. The practical
feasibility will still depend on the mode density, rotational
splitting, and photometric precision of a given event, but the
time-dependence of the amplitude ratios during the transit potentially offers a
qualitatively new mode-identification diagnostic.

\section{Conclusions}
\label{sec:conclusion}

We studied how gravitational microlensing changes the disk-integrated visibility of stellar oscillations. Our main results are as follows.

First, close microlensing events can lift the geometric cancellation that normally suppresses high-degree modes in unresolved stars. When the projected source-lens separation is comparable to the angular radius of the source, the differential magnification across the disk increases the visibility of high-degree modes by up to three orders of magnitude in our simulations, making envelope-sensitive modes potentially detectable.

Second, the non-axisymmetric magnification pattern redistributes power among the azimuthal components of rotationally split multiplets. In standard asteroseismology, this distribution constrains only the inclination of the spin axis; during a close microlensing event, it also depends on the sky-projected azimuth, imprinting the three-dimensional spin orientation onto the power spectrum. In an illustrative synthetic case, measurements of the relative powers for $l=1,2,3$ at three lens-source separations recovered the full spin orientation with uncertainties of $\sigma_i=2.9^\circ$ and $\sigma_\gamma=11.5^\circ$.

Third, the method is limited by event rarity and duration, and by the difficulty of identifying modes in the more crowded spectrum. The favourable alignment lasts only while the lens passes within a few source radii, and the expected number of bright, high-quality events is small. Red giants are the most promising targets because their large radii increase both the event probability and the time available for seismic measurements.

Microlensing therefore offers a narrow but distinctive route to spatially selective asteroseismology beyond the Sun. A single successful phakoseismic event would provide access to envelope-sensitive oscillation modes, and possibly the full spin-axis geometry, for a distant star that would otherwise remain unresolved.

\begin{acknowledgements}
This work was supported by a research grant (VIL70041) from VILLUM FONDEN. We thank Jørgen Christensen-Dalsgaard for comments on an early draft of the paper and for providing the ADIPLS oscillation frequencies used in the power spectra.
\end{acknowledgements}

\bibliographystyle{aa}
\bibliography{sn-bibliography}

\end{document}